\documentclass[aps,prl,twocolumn,showpacs,superscriptaddress,floatfix]{revtex4-1}

\usepackage{amsmath}
\usepackage{amssymb}
\usepackage{graphicx, hyperref}
\usepackage{grffile} % to handle dots in graphics filenames
\usepackage{float}
\usepackage{color}
\usepackage{subfigure}
\usepackage{soul}
\hypersetup{draft}
\usepackage{lipsum}

\newcommand{\tb}[1]{\textbf{#1}} 
 
\newcommand{\mb}[1]{\mathbf{#1}}

\newcommand{\kk}{\mathbf{k}}
\newcommand{\qq}{\mathbf{q}}

\newcommand{\unitE}{$[V/\mathrm{a_0}]$ }
\newcommand{\unitW}{$\mathrm{eV}$ }

\begin{document}

\author{Ankit~Kumar}
\email{akumar13@ncsu.edu}
\affiliation{Department of Physics, North Carolina State University, Raleigh, North Carolina 27695, USA}

\author{S.~Johnston}
\affiliation{Department of Physics and Astronomy, The University of Tennessee, Knoxville, Tennessee 37996, USA}

\author{A.~F.~Kemper}
\email{akemper@ncsu.edu}
\affiliation{Department of Physics, North Carolina State University, Raleigh, North Carolina 27695, USA}

%\title{Effect of forward scattering on the superconducting state: A time-domain study}
\title{Identifying a forward scattering superconductor through pump-probe spectroscopy}

 \begin{abstract}
 Electron-boson scattering that is peaked in the forward direction has been suggested as an essential ingredient for
 enhanced superconductivity observed in FeSe monolayers. Here, we study the superconducting state of a system dominated
 by forward scattering in the time-domain and contrast its behavior against the standard isotropic BCS case for both
 $s$- and $d$-wave symmetries. An analysis of the electron's dynamics in the pump-driven non-equilibrium state reveals
 that the superconducting order in the forward-focused case is robust and persistent against the pump-induced
 perturbations. The superconducting order parameter also exhibits a non-uniform melting in momentum space. We show that
 this behavior is in sharp contrast to the isotropic interaction case and propose that time-resolved approaches are a
 potentially powerful tool to differentiate the nature of the dominant coupling in correlated materials.
 \end{abstract}

 \maketitle
 The discovery of enhanced superconductivity in FeSe intercalates \cite{Liu2012} and ultra thin films of FeSe grown on
 oxides substrates \cite{Qing-Yan2012}, has attracted considerable interest. In these systems, significant electron
 doping drives the $\Gamma$-centered hole bands below the Fermi level $E_\mathrm{F}$ \cite{2013_He}. At face value, this
 Fermi surface topology challenges the picture of $s_\pm$ pairing mediated by spin fluctuations enhanced by Fermi
 surface nesting \cite{Mazin2008}. Motivated by this, several alternative proposals have been advanced, including
 contributions from the so-called incipient hole band below $E_\mathrm{F}$ \cite{Mishra2016, Linscheid2016, Chen2015};
 or from some other bosonic mode such as nematic fluctuations \cite{Kang2016} or phonons \cite{Qing-Yan2012, Coh2015,
 2014Lee, 2015Lee, Rademaker2016, Choi2017}. In particular, a unique cross interfacial electron-phonon ($e$-ph) coupling
 has been invoked to explain the higher T$_c$ values observed in FeSe films grown on oxide substrates in comparison to
 the doped FeSe intercalates \cite{2014Lee, 2015Lee, Li2016, Rebec2017}; this interpretation, however, is controversial
 \cite{2017Zhou, 2017Nekrasov, 2017Jandke, 2017Song, 2018Li, 2018Zhang}, and finding experimental probes that might provide new
 insight into this problem is imperative. 

 A common feature of the scenarios based on nematic fluctuations or cross-interface $e$-ph coupling is the importance of
 forward scattering, where the relevant electron-boson coupling constant is peaked strongly at small momentum transfers
 ${\bf q}$.  Such a momentum structure can have significant consequences for superconductivity
 \cite{Kulic2004,Rademaker2016, Roof2016,Kulic2017}, enhancing pre-existing superconductivity in unconventional pairing
 channels, or resulting in significant $T_c$ values if the interaction is sufficiently peaked around ${\bf q} = 0$
 \cite{Bulut1996, Aperis2011, Kulic2004, 2014Lee, Rademaker2016, Kulic2017}. These results naturally raise the question: how
 can one uniquely identify superconductivity mediated by a forward-focused interaction? 

 %Fig01
 \begin{figure}
 %\fbox{
        % fbox { trim left bottom right top }
 \includegraphics[width=0.49\textwidth, trim=0.4cm 0.8cm 0.76cm 1.1cm, clip=true]{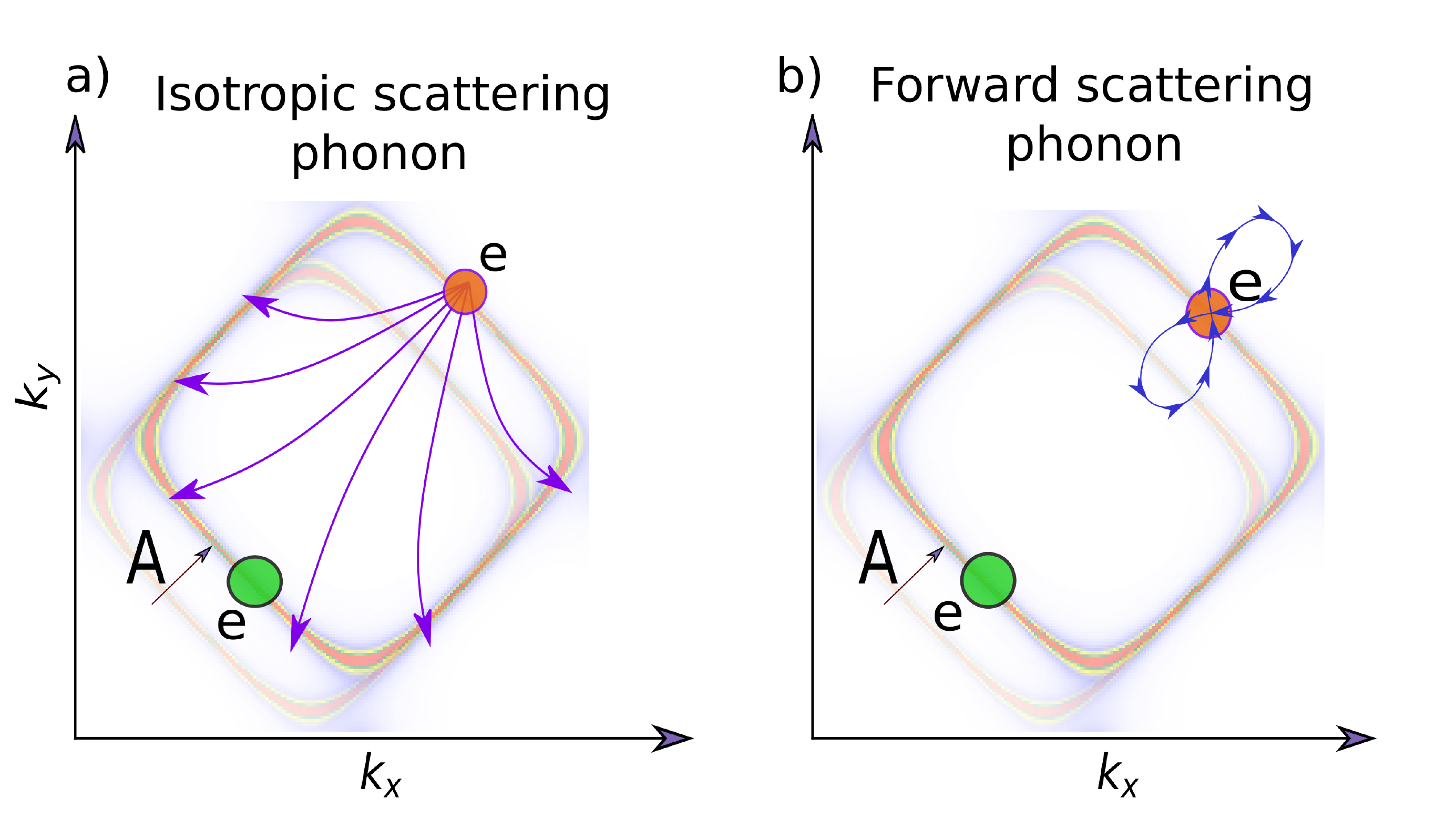}
 %}
    \caption{(Color online) An illustration of the allowed scattering processes in the system, where the Fermi surface
    is displaced by a pump $\mathbf{A}$(t) applied along the (11) direction. \tb{a)} Scattering of electrons by
    isotropically coupled optical phonons. The electron may scatter to all allowed $\kk$ states. \tb{b)} Scattering of
    electrons by optical phonons with a bias towards the forward scattering. The electron has restricted phase space
    available to scatter to other states.}
    \label{fig:illustration}
 \end{figure}

 We address this issue in this Letter by studying the problem in the time-domain. We consider superconductivity mediated
 by an optical phonon in the idealized perfect forward scattering limit (i.e. a delta-function coupling) and contrast
 its behavior to the case where superconductivity is mediated by an isotropic interaction. (While we adopt a phonon mode
 for simplicity, our results are also applicable to the pairing mediated by nematic fluctuations.) In both cases, we
 drive the system into a non-equilibrium state by an ultrafast pump pulse and study the properties of the transient
 non-equilibrium state. We find that the two interactions show significant contrasting behavior in spectroscopic
 measurements. First, the suppression of the superconducting order in the isotropic case shows a strong dependence on
 the pump fluence. The forward scattering superconducting state is comparatively robust against pump-induced
 suppression. The energy absorption is also much smaller in the forward focused case. Second, while the suppression of
 the superconducting gap in the isotropic case is uniform in momentum space, the forward scattering shows a specific
 momentum-dependent suppression of the gap. This particular contrasting behavior can be measured in momentum-
 and time-resolved measurements, such as trARPES, and can provide evidences to either validate or falsify the forward
 scattering mechanism in FeSe monolayers. 

 % FIG02 -  ARPES
 \begin{figure*}
    %\fbox{
        % fbox { trim left bottom right top }
        \includegraphics[width=0.98\textwidth, trim=0.26cm 0.29cm 0.3cm 0.32cm, clip=true]{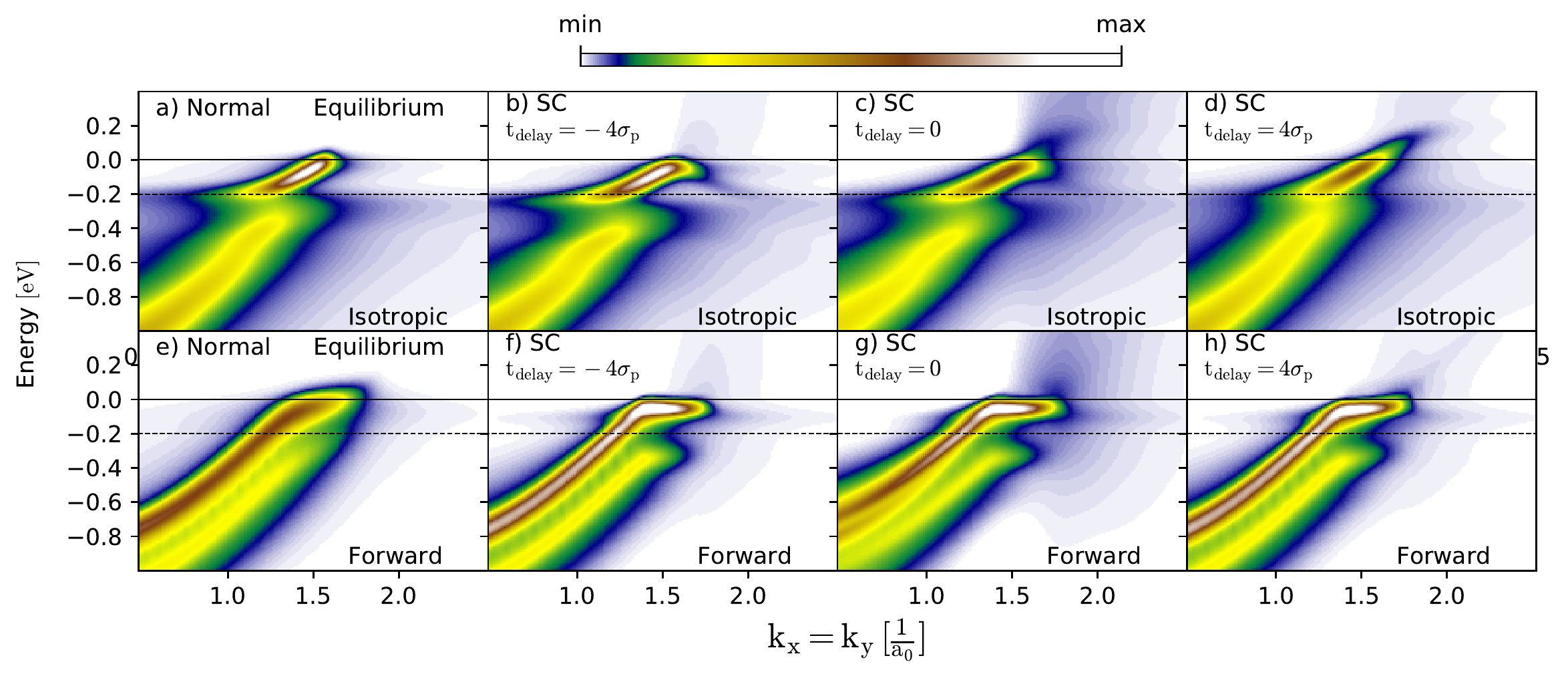}
     %}
        \caption{(Color online) tr-ARPES spectra of the isotropic scattering and the forward scattering along the (1,1)
        direction for $s$-wave pairing symmetry. Panels \tb{a} and \tb{e} display the equilibrium normal state. The Fermi
        energy ($E_\mathrm{F}$) and phonon frequency are shown by the solid horizontal line and dashed line respectively.
        Panels \tb{b}-\tb{d} and \tb{f}-\tb{h} show the time-resolved ARPES spectra of the superconducting state when a
        pump pulse drives the system out of
        equilibrium.}
        \label{fig:arpes}
 \end{figure*}
 
 The observed effects can be understood as a direct consequence of the \textit{restricted phase space}
 available to scattering process in the forward scattering case. Figure~\ref{fig:illustration} shows schematic scattering
 processes at $E_\mathrm{F}$, where electrons are driven by an ultrafast pump field $\mb{A}(t)$. In the case of an
 isotropic $e$-ph interaction, an electron can scatter to all energy and momentum conserving eigenstates. In contrast, in the forward
 focused case, the only final states close to the initial state are available to an electron, i.e. momentum transfer
 $\qq\approx0$. This restriction on phase space hinders the energy absorption in forward scattering and enables
 non-trivial momentum structure of the gap \cite{Wang2016}, resulting in the observed effects. 

  % \section{Parameters}
 We consider the Holstein Hamiltonian on 2D lattice 
 \begin{align}
 \mathcal H &= \sum_{\kk,\sigma}\xi(\kk) c^\dagger_{\kk,\sigma} c^{\phantom\dagger}_{\kk, \sigma} 
         + \sum_{\qq,\nu} \Omega_\nu  \left(b_{\qq,\nu}^\dagger b^{\phantom\dagger}_{\qq,\nu} + \frac{1}{2} \right) \nonumber\\
   &+ \frac{1}{\sqrt{N}}\sum_{\substack{\nu,\sigma\\\kk,\qq}}g(\kk,\qq) c_{\kk+\qq, \sigma}^\dagger
         c^{\phantom\dagger}_{\kk,\sigma} \left( b^{\phantom\dagger}_{\qq,\nu} +
         b_{-\qq,\nu}^\dagger \right). 
 \end{align} Here, $\xi(\kk)$ nearest neighbor tight-binding energy dispersion measured relative to the chemical
 potential, $c^\dagger_\kk, c^{\phantom\dagger}_\kk$ ($b^\dagger_\qq, b_\qq)$ are the standard creation and annihilation
 operators for an electron (phonon), $g(\kk,\qq)$ is the momentum-dependent $e$-ph coupling constant, and $\Omega_\nu$
 is the frequency for the Einstein phonon mode $\nu$. Here we take an effective-minimal model to capture the distinctive
 features of the forward scattering. Although FeSe monolayers are quite complicated (e.g. in their multiband nature), we
 assume the forward scattering mechanism to be dominant and ignore further complications. This is sufficient to test
 the forward scattering mechanism. 

 The superconducting state is modeled by the standard two-particle Nambu spinors in the self-consistent
 Migdal-Eliashberg formalism and time evolution is done by solving the Gor'kov equations self-consistently on the
 Keldysh contour \cite{Kemper2015} with an ultrafast pump field, included via Peierls' substitution.
 We consider two types of $e$-ph interactions: an isotropic coupling with $g(\kk,\qq) = g$, a constant, and a
 \textit{forward scattering} interaction where $g(\kk,\qq)$ is taken in the limit of perfect forward scattering
 $g(\kk,\qq) = g\delta_{\qq,\mathbf{0}}$ \cite{Wang2016} to reduce the computational cost. The different coupling
 strengths are tuned to get approximately same superconducting gap ($\Delta \sim 51$ meV) for both isotropic and forward
 focused interactions (see Supplemental Material). For completeness, we include a solution in the $d$-wave channel as
 well.
 
 %\section{Results and discussion}
 \begin{table}[t]
 \caption{Parameters used in the simulation}
 \label{tab:parameters}
 \begin{ruledtabular}
 \begin{tabular}{lcr}
 Parameters & Isotropic & Forward \\
 \hline
 Phonon frequency ($\Omega_{1,2}$) & $ 0.2, 0.001$ \unitW & $0.2, 0.05$ \unitW \\
 Phonon coupling ($g^2_{1,2}$) &$0.12, 0.001$ \unitW & $0.034, 0.01$ \unitW\\
 \hline
 Band parameters &\multicolumn{2}{c}{ $V_{nn}=0.25 \, \mathrm{eV}$, $\mu = 0.0 \, \mathrm{eV}$} \\
 Temperature &\multicolumn{2}{c}{ $\approx 83 \, \mathrm{K}$ } \\
 %Energy dispersion &\multicolumn{2}{c} { $-2V_{nn}(cos(k_x) + cos(k_y)) $} \\
 Pump pulse &\multicolumn{2}{c} { $\omega_p = 1.5\,\mathrm{eV}$ , $\sigma_p = 8 \; \mathrm{[1/eV]}$ } \\
 Probe pulse &\multicolumn{2}{c} { $\sigma = 25 \; \mathrm{[1/eV]}$ } \\
 \end{tabular}
 \end{ruledtabular}
 \end{table}

 The nature of the \textit{perfect} forward scattering restricts the system's ability to absorb any energy via scattering
 processes. Realistically, the coupling has some finite width \cite{Zhou2016,2014Lee,Kulic2017}, and there are
 additional phonons to scatter from in the system. To account for these effects, we include a second isotropically
 coupled phonon in the system, whose coupling is weak enough that it can dissipate energy but does not significantly
 contribute to the superconductivity. Table~\ref{tab:parameters} lists the parameters used in the simulation. 
 
 First, we compare the effect of the pump pulse on the superconductor through the
 single-particle electron spectral function. Figure~\ref{fig:arpes} displays the calculated time-resolved and
 angle-resolved photoemission spectroscopy (tr-ARPES) spectra of an $s$-wave superconductor for both types of scattering.
 For isotropic scattering (the top panels) the equilibrium band structure in normal state (Fig.~\ref{fig:arpes}\tb{a})
 shows a kink at the phonon energy ($\pm \Omega_1$), which is common for strong coupling to phonon in the system
 \cite{Engelsberg1963}. The normal state band structure for the forward focused case (Fig.~\ref{fig:arpes}\tb{e})
 exhibits different behavior \cite{Rademaker2016}; most notably there is a replica band below the main band, similar to
 the experimentally observed \cite{2014Lee, Peng2014, Rebec2017} features of the FeSe monolayer superconductors that
 have been explained by forward scattering from phonons. The band is flatter in the vicinity of $E_\mathrm{F}$ due to the perfect forward scattering limit \cite{Wang2016}. 

 %Fig03 - Gap supression. 
 \begin{figure}
    \center
 %   \fbox{
        % fbox { trim left bottom right top }
        \includegraphics[width=0.49\textwidth,clip=true, trim=0.1cm 0.2cm 0.2cm 0.1cm]{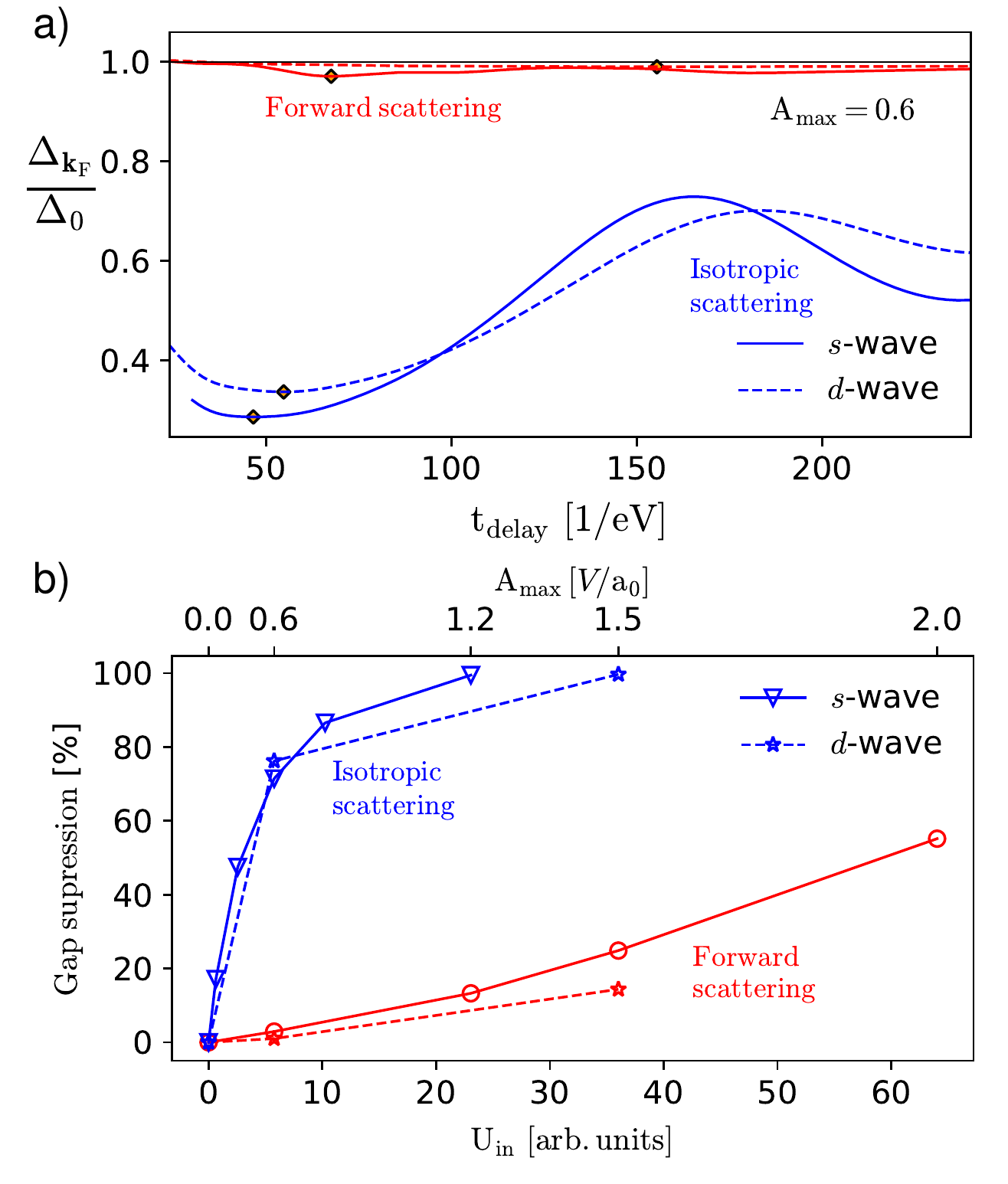}
 %   }
    \caption{(Color online) The effect of the pump on the superconducting order parameter. Panel \tb{a} displays the
    maximum order parameter (at the Fermi momentum $\kk=(\pi/2,\pi/2)$ for $s$-wave, and at $\kk=(\pi,0)$ for
    $d$-wave) as a function of delay time for fluence $\mathrm{A_{max}} = 0.6$ \unitE. The $s$-wave and $d$-wave pairing
    symmetries are compared for both type of scattering. The order parameter is normalized by the equilibrium value
    $\Delta_0$ ($\sim 51$ meV).  The diamond marker indicates the maximum suppression of the superconducting order.
    Panel \tb{b} shows the maximum suppression of the superconducting order as a function of incident energy (and
    fluence top X-axis).}
    \label{fig:compare-gap-time}
 \end{figure}
 
 The remaining panels in Fig.~\ref{fig:arpes} show the tr-ARPES spectra in the superconducting state, where a pump pulse
 of amplitude $\mathrm{A_{max}} = 1.2$ \unitE is used to drive the system along the ($11$) direction. This fluence is high
 enough to melt the superconducting order significantly for isotropic pairing \cite{Kemper2015}, as shown in
 Fig.~\ref{fig:arpes} \tb{c,d}. The superconducting state in equilibrium is shown in Fig.~\ref{fig:arpes}\tb{b}.
 The spectra show the characteristics of a strong-coupling superconductor: a
 decrease of spectral weight near $E_\mathrm{F}$ indicating the opening of a gap, the appearance of a
 particle-hole symmetric shadow band, and a shift of the kink by the maximum gap value \cite{Eschrig_2003, lee_2008}.
 When the pulse hits the system at $\mathrm{t_{delay}}=0$ $\mathrm{[1/eV]}$ (Fig.~\ref{fig:arpes}\tb{c}), the spectral
 weight redistributes itself along ($11$) direction. The movement of spectral weight towards $E_\mathrm{F}$ represents
 the melting of the superconducting order. After the pump
 (Fig.~\ref{fig:arpes}\tb{d}), we observe that the superconducting order has partially melted, as the spectral weight is
 enhanced near $E_\mathrm{F}$. The spectrum suggests that the system has entered the normal state; however,
 superconductivity is still present \cite{Kemper2015}, as we will show in the next section. 
      \begin{figure*}
 %     \fbox{
        % fbox { trim left bottom right top }
      \includegraphics[width=0.98\textwidth, trim=0.3cm 0.2cm 0.22cm 0.2cm, clip=true]{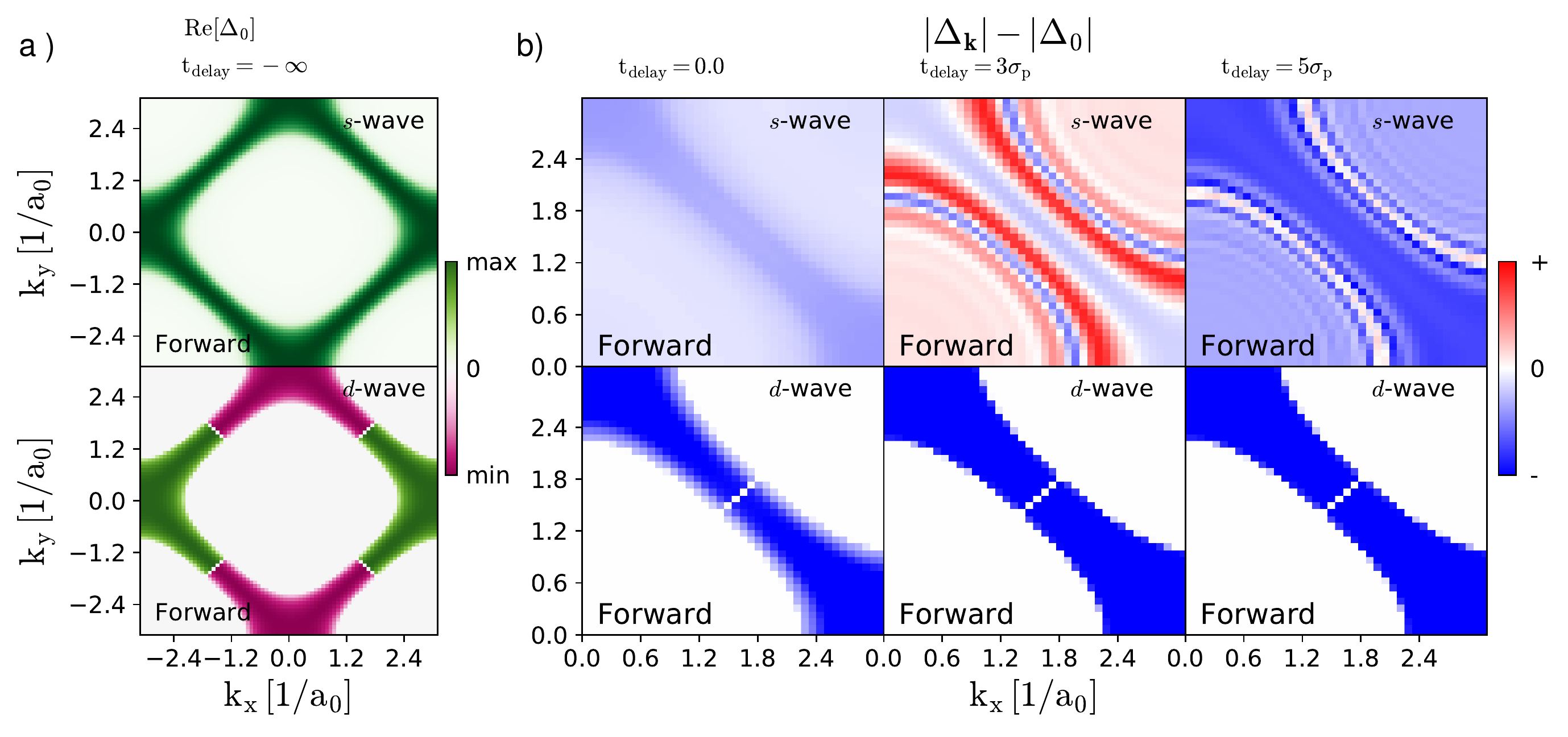}
 %     }
         \caption{(color online) The momentum resolved dynamics of the $s,d$-wave superconducting order for 
         the forward scattering. Panel \tb{a} displays the equilibrium superconducting state. Panel \tb{b} shows the change
         in magnitude of the superconducting order parameter at three delay times $(0, 3\sigma_p, 5\sigma_p)$ when a
         pump drives the system out of
         equilibrium.
         } \label{fig:k-gap}
     \end{figure*}
 
 Figure~\ref{fig:arpes}\tb{f} shows the contrasting case of the superconducting state in the forward scattering
 scenario. We observe a flattening of the band near $E_\mathrm{F}$ compared to the normal state
 (Fig.~\ref{fig:arpes}\tb{e}). Figs.~\ref{fig:arpes}\tb{g} and~\ref{fig:arpes}\tb{h} show the spectrum when the same
 pump pulse as used in the isotropic scattering case drives the system. We observe a weaker redistribution of the
 spectral weight, even for times after the pump (Fig.~\ref{fig:arpes}\tb{h}), indicating that the superconducting
 order based on forward-focused pairing is remarkably robust and persistent. 

 %This result can be understood via the very nature of the forward scattering: as the interaction is peaked with
 %$\mb{q\approx0}$, electrons have a very restricted phase space around the initial state to scatter to
 %(Fig.~\ref{fig:illustration}). Consequently, when a pump drives the electrons, the system absorbs much less energy
 %compared to the isotropic scattering case and the superconducting order largely persists in the system. A calculation
 %of the energy absorption may be found in the Supplement.

 To gain further insight we examine the evolution of the spectral gap in the time-domain, which is estimated
 from the retarded anomalous self energy $\Sigma_F^{R}(\kk;t,t')$
 \begin{align}
 \Delta_\kk(\mathrm{t_{delay}, \omega_{rel}=0)  = \int dt_{rel} \Sigma_F^{R}(
 \kk;t_{delay},t_{rel}}), 
 \end{align}
 where a Wigner transformation $\mathrm{t_{delay} = \frac{t+t'}{2}}$, $\mathrm{t_{rel} = t-t'}$ is used to transform
 the reference frame to average and relative times. First we calculate the order parameter at the Fermi momentum
 $\Delta( \mathrm{t_{delay}} ) \equiv \Delta_{\kk = \kk_\mathrm{F}}(\mathrm{t_{delay}})$ for both types of scattering;
  see Fig.~\ref{fig:compare-gap-time}\tb{a} which shows $\Delta(\mathrm{t_{delay}})$
  for different pairing symmetries after normalizing by the equilibrium value $\Delta_0$. We
 observe that the forward scattering $\Delta$ melts less than the isotropic scattering $\Delta$.
 To quantify the suppression we determine the minimum value of the transient order parameter and plot it as
 function of fluence and absorbed energy in Fig.~\ref{fig:compare-gap-time}\tb{b} (a calculation of the energy
 absorption may be found in the Supplement). Since less energy is absorbed when forward scattering is the dominant
 source of scattering, a superconducting order built from forward scattering pairing will also be affected less. Also,
 we observe that after the pump $\Delta$ relaxes in an oscillatory fashion (Fig.~\ref{fig:compare-gap-time}\tb{a}).
 These amplitude mode, or Higgs mode oscillations have been proposed and seen before in literature
 \cite{Kemper2015,Nosarzewski2017,Matsunaga_2013,Matsunaga_2014, Murakami_2016} for pumped superconductors. In forward
 scattering, we observe similar oscillations for low fluences. 
 
 %Fig06  - Time traces.
  \begin{figure}
    \center
     %   \fbox{
        % fbox { trim left bottom right top }
        \includegraphics[width=0.49\textwidth, clip=true, trim=0.2cm 0.2cm 0.2cm 0.1cm]{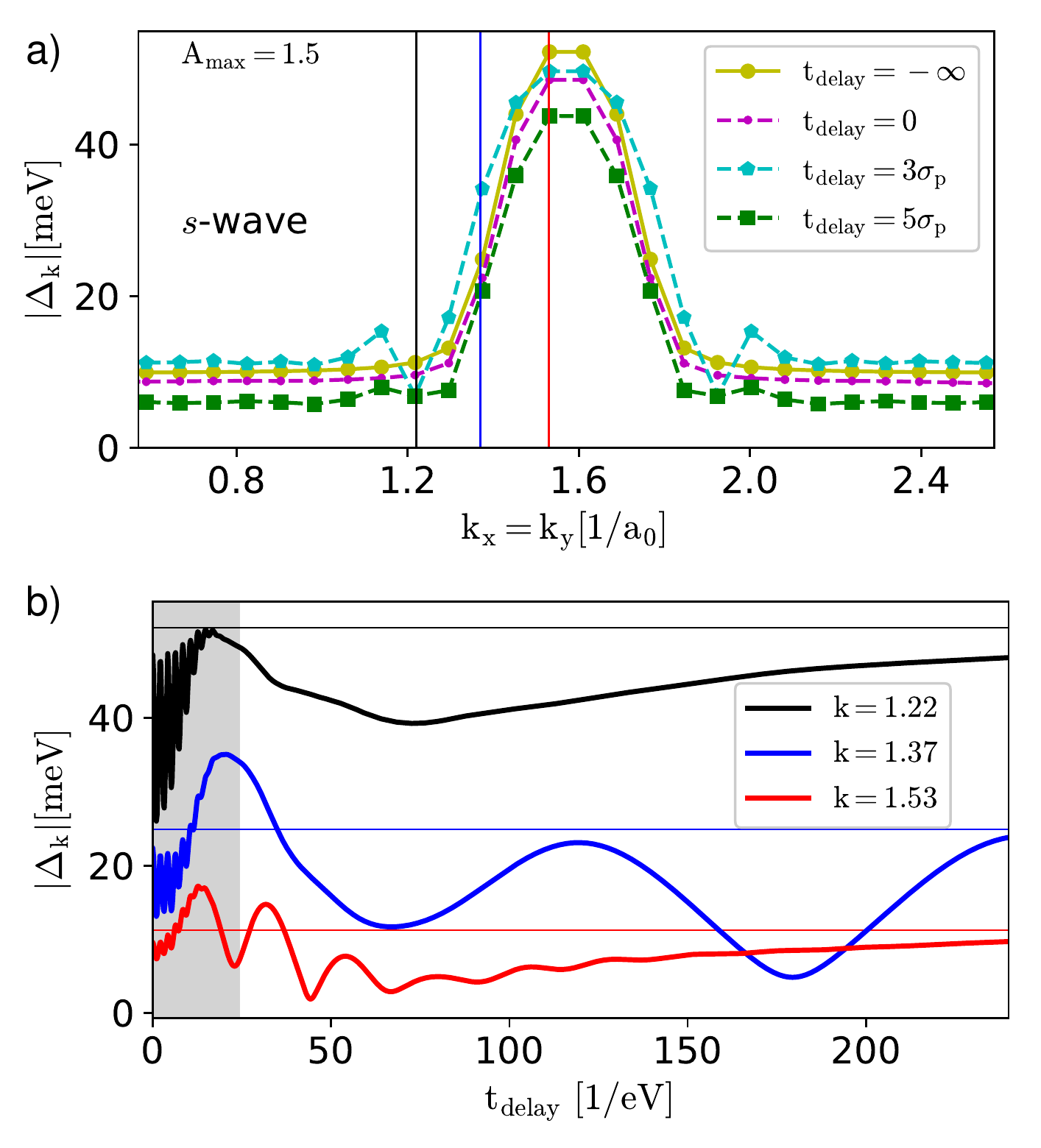}
     %   }
        \caption{(Color online) Panel \tb{a} shows magnitude of the $s$-wave superconducting order along the symmetry
        line $\mathrm{k_x=k_y}$ for different delay times. Panel \tb{b} displays magnitude of the order parameter at
        three fixed $\mb{k}$ points, marked by the vertical straight lines in panel \tb{a}, as a function of delay time.
        Horizontal straight lines in panel \tb{b} are the equilibrium value for each momentum point, respectively.  }
        \label{fig:time-traces}
  \end{figure}

  Finally, we analyze the momentum structure of the order parameter $\Delta_\kk(\mathrm{t_{delay}})$.
  Isotropic scattering pairing and forward scattering pairing result in \textit{momentum independent} and
  \textit{momentum dependent} $s/d$-wave order parameters, respectively \cite{Rademaker2016, Wang2016}. 
  By momentum dependence, we mean on top of their respective $s/d$ symmetries.
  We study the evolution of the forward scattering driven momentum dependent
  gap for $s$-wave and $d$-wave pairing. Figure~\ref{fig:k-gap}\tb{a} shows the magnitude of $\Delta_\kk$ in equilibrium.
  Although the order parameter is $s/d$-wave ($\mathrm{A_{1g}/B_{1g}}$), it is not uniform across the Brillouin zone due to the
  forward focus in the interaction \cite{Wang2016}. Figure~\ref{fig:k-gap}\tb{b} shows the
  change in the transient order parameter when the system is driven out of equilibrium
  by a pump of amplitude $\mathrm{A_{max}} = 1.5$ $\mathrm{[V/a_0]}$. 
  %We use a high pump field to enhance the suppression of the order parameter quantitatively.
  
  For $s$-wave pairing, $\Delta_\kk$ is suppressed the most at $E_\mathrm{F}$ (third snapshot in
  Fig.~\ref{fig:k-gap}\tb{b} at $\mathrm{t_{delay}} = 5\sigma_p$). Surprisingly, in some regions of the Brillouin zone,
  the order parameter increases when the pump is active (second snapshot in Fig.~\ref{fig:k-gap}\tb{b} at
  $\mathrm{t_{delay}} = 3\sigma_p$). However, overall change in the order parameter is negative because the system
  absorbs finite amount of energy which leads to dissociation of the Cooper pairs. For $d$-wave pairing, the
  superconducting gap does not show any enhancement after the pump and is suppressed at all $\mathbf{k}$ points. To
  analyze the dynamics further we show the order parameter for $s$-wave gap as a function of $\kk$ along the ($11$)
  direction for different delay times in Fig.~\ref{fig:time-traces}\tb{a}. Figure~\ref{fig:time-traces}\tb{b} displays
  magnitude of the order parameter as a function of delay time for three representative $\kk$ points marked by the
  vertical lines in Fig.~\ref{fig:time-traces}\tb{a}, which highlights the nonuniform changes in $\Delta_\kk$. Such
  order suppression is characteristically different in isotropic interaction scenario, where the suppression is larger
  and momentum independent, shown in Fig.~\ref{fig:compare-gap-time}.

 %\section{conclusion}
 In summary, we have studied the effect of $e$-ph scattering focused in the forward direction on the superconducting
 state in the time-domain and compared our results with isotropic scattering case. The forward scattering superconductor
 shows a robust, persistent superconducting order against the pump-induced perturbations. We also observe a momentum
 structure of the superconducting order parameter in the forward scattering. The order parameter exhibits a non-uniform
 melting in momentum space in the non-equilibrium state. These characteristic features of the forward scattering are in direct
 contrast with the conventional BCS isotropic superconductors, where a reasonable fluence can melt the superconducting
 order and the suppression of the superconducting oder shows no momentum dependence. In fact, even unconventional (cuprate)
 superconductors show a rapid suppression of the gap for weak pump pulse \cite{2011_Graf}. We explain our results for
 the forward pairing superconducting order using the restriction of phase space available for the scattering process to
 occur, which results in less absorption of energy. We support our explanation by estimating the absorbed energy and
 studying the dynamics of the superconducting order parameter in the time domain. These distinctive features of forward
 scattering may be detectable using momentum-resolved experiments such as time-resolved ARPES. 
 
 Recent work has suggested a weak forward scattering phonon as a source of the high T$_C$ as an enhancement of an
 unconventional pairing mechanism.  In this scenario, the time-resolved signatures discussed here would appear
 essentially identical to a standard BCS superconductor since the lack of phase space does not apply, and a quantitative
 comparison would have to be made with respect to some reference material; an FeSe intercalate which has no interfacial
 phonons may be used for an absolute comparison to see if there is an increase in energy absorption commensurate with
 the enhanced pairing.

 \section{ACKNOWLEDGMENTS} We acknowledge fruitful discussion with Doug Scalapino, Yan Wang, Louk Rademaker, Avinash
 Rustagi and Omadillo Abdurazakov. This research used resources of the National Energy Research Scientific Computing
 Center, a DOE Office of Science User Facility supported by the Office of Science of the U.S. Department of Energy under
 Contract No. DE-AC02-05CH11231.
 
% \bibliography{refs.bib}

\end{document}